\documentclass[aps,prb,showpacs,groupedaddress,amssymb,twocolumn]{revtex4}
\usepackage{bm,amsmath,graphicx}
\begin{document}

\title{Infrared catastrophe and tunneling into strongly correlated electron systems: Perturbative x-ray edge limit}
\author{Kelly R. Patton and Michael R. Geller}
\affiliation{Department of Physics and Astronomy, University of Georgia,  Athens, Georgia 30602-2451}

\date{March 31, 2005}

\begin{abstract}
The tunneling density of states exhibits anomalies (cusps, algebraic suppressions, and pseudogaps) at the Fermi energy in a wide variety of low-dimensional and strongly correlated electron systems. We argue that in many cases these spectral anomalies are caused by an infrared catastrophe in the screening response to the sudden introduction of a new electron into the system during a tunneling event. A nonperturbative functional-integral method is introduced to account for this effect, making use of methods developed for the x-ray edge singularity problem. The formalism is applicable to lattice or continuum models of any dimensionality, with or without translational invariance. An approximate version of the technique is applied to the 1D electron gas and the 2D Hall fluid, yielding qualitatively correct results.
\end{abstract}

\maketitle
\clearpage

\section{TUNNELING AND THE INFRARED CATASTROPHE}\label{introduction section}

The tunneling density of states (DOS) is known to exhibit spectral anomalies such as cusps, algebraic suppressions, and pseudogaps at the Fermi energy in a wide variety of low-dimensional and strongly correlated electron systems, including
\begin{itemize}

\item[(i)] all 1D metals;\cite{TomonagaPTP50,MattisJMP65,HaldaneJPC81,HaldanePRL81,KanePRB92}

\item[(ii)] the 2D diffusive metal;\cite{AltshulerPRL80,Altshuler85,LeeRMP85,Levitov97,KamenevPRB99,ChamonPRB99}

\item[(iii)] the 2D Hall fluid;\cite{EisensteinPRL92,YangPRL93,HatsugaiPRL71,HePRL93,JohanssonPRL93,KimPRB94,AleinerPRL95,HaussmannPRB96,LeonardJPCM98,WangPRL99}

\item[(iv)] the edge of the confined Hall fluid.\cite{WenPRL90,WenPRB90a,WenPRB91a,MoonPRL93,KanePRL94,KanePRB95,MillikinSSC96,ChangPRL96,HanPRB97a,HanPRB97b,Conti97,GraysonPRL98,ShytovPRL98,ContiJPCM98,LopezPRB99,ZulickePRB99,PruiskenPRB99,KhveshchenkoSSC99,MoorePRB00,AlekseevJETP00,ChangPRL01,GoldmanPRL01,HilkePRL01,LopezPRB01,PasquierPRB01,TsiperPRB01,LevitovPRB01,MandalSSC01,WanPRL02,MandalPRL02,ZulickePRB02,HuberPRL05}

\end{itemize}
In this paper we propose a unified explanation for these anomalies, develop a nonperturbative functional-integral formalism for calculating the electron propagator that captures the essential low-energy physics, and apply a simplified approximate version of the method to cases (i) and (iii), yielding qualitatively correct results.

We claim that the physical origin of the DOS anomalies in the above systems is the infrared catastrophe of the host electron gas caused by the sudden perturbation produced by an electron when added to the system during a tunneling event. This infrared catastrophe, a singular screening response of a conductor to a localized potential turned on abruptly in time caused by the large number of electron-hole pairs made available by the presence of a sharp Fermi surface, is known to be responsible for the singular x-ray optical and photoemission spectra of metals,\cite{MahanPR67,Nozieres&DeDominicisPR69,OhtakaRMP90} Anderson's related orthogonality catastrophe,\cite{AndersonPRL67,HamannPRL71} and the Kondo effect.\cite{AndersonPRL69,YuvalPRB90} To understand the connection to tunneling, imagine the tunneling electron being replaced by a negatively charged, distinguishable particle with mass $M$. In the $M \! \rightarrow \! \infty$ limit, the potential produced by the tunneling particle is identical---up to a sign---to the abruptly turned-on hole potential of the x-ray edge problem, and an infrared catastrophe of the host electrons would be expected. 

Tunneling of a real, finite-mass electron is different because it recoils, softening the potential produced. However, in the four cases listed above, there is some dynamical effect that suppresses recoil and produces a potential similar to that of the infinite-mass limit: In case (i) the dimensionality of the system makes charge relaxation slow; In case (ii), the disorder suppresses charge relaxation; In case (iii),  the Lorentz force keeps the injected charge localized; and case (iv) is a combination of cases (i) and (iii). It is therefore reasonable to expect remnants of the infinite-mass behavior. The real electron is also an indistinguishable fermion, unable to tunnel into the occupied states below $\epsilon_{\rm F}$.
 
Our analysis proceeds as follows: We use a Hubbard-Stratonovich transformation to obtain an exact functional-integral representation for the imaginary-time Green's function 
\begin{equation}
G({\bf r}_{\rm f}\sigma_{\rm f},{\bf r}_{\rm i} \sigma_{\rm i},\tau_0) \equiv - \big\langle T \psi_{\sigma_{\rm f}}({\bf r}_{\rm f} , \tau_0) 
{\bar \psi}_{\sigma_{\rm i}}({\bf r}_{\rm i},0) \big\rangle_{\! \scriptscriptstyle H},
\label{G definition}
\end{equation}
and identify a ``dangerous" scalar field configuration $\phi_{\rm xr}({\bf r},\tau)$ that causes an infrared catastrophe. Assuming $\tau_0 \ge 0$ and an electron-electron interaction potential $U({\bf r})$, 
\begin{equation}
\phi_{\rm xr}({\bf r},\tau) \equiv U\big[{\bf r} - {\bf R}(\tau)\big] \Theta(\tau) \Theta(\tau_0-\tau),
\label{general phi xr}
\end{equation}
where ${\bf R}(\tau) \equiv {\bf r}_{\rm i} + (\frac{ {\bf r}_{\rm f} - {\bf r}_{\rm i}}{\tau_0}) \tau$ is the straight-line trajectory connecting ${\bf r}_{\rm i}$ to ${\bf r}_{\rm f}$ with velocity $({\bf r}_{\rm f} - {\bf r}_{\rm i})/\tau_0.$ For the case of the tunneling DOS at point ${\bf r}_0,$ we have ${\bf r}_{\rm i} = {\bf r}_{\rm f} = {\bf r}_0$ and 
\begin{equation}
\phi_{\rm xr}({\bf r},\tau) = U({\bf r} - {\bf r}_0) \Theta(\tau) \Theta(\tau_0-\tau),
\label{DOS phi xr}
\end{equation}
which is the potential that would be produced by the added particle in (\ref{G definition}) if it had an infinite mass. Fluctuations about $\phi_{\rm xr}$ account for the recoil of the finite-mass tunneling electron. 

In this paper we introduce an approximation where the fluctuations about $\phi_{\rm xr}$ are entirely neglected, which we shall refer to as the x-ray edge limit. We expect the x-ray edge limit to give qualitatively correct results in the systems (i) through (iv) listed above; the effect of fluctuations will be addressed in future work. The x-ray edge limit of our formalism can be implemented exactly, without further approximation, only for a few specific models. Here we will instead carry out an approximate but more generally applicable analysis of the x-ray edge limit by resumming a divergent perturbation series caused by the infrared catastrophe. We apply this ``perturbative" x-ray edge method to the 1D electron gas and 2D spin-polarized Hall fluid, both having a short-range interaction of the form
\begin{equation} 
U({\bf r}) = \lambda \delta({\bf r}).
\label{short range interaction definition}
\end{equation}
The exact implementation of the x-ray edge limit for these two cases will be presented elsewhere.

\section{INTERACTING PROPAGATOR}\label{functional integral section}

We consider a $D$-dimensional interacting electron system, possibly in an external magnetic field. The grand-canonical Hamiltonian is
\begin{eqnarray*}
&H& = \sum_\sigma \! \int \! d^{\scriptscriptstyle D}r \, \psi^\dagger_\sigma({\bf r}) \, \bigg[{\Pi^2 \over 2m} + v_0({\bf r}) - \mu_0 \bigg] \psi_\sigma
({\bf r}) \nonumber \\
{\hskip -0.4in} &+& \! {\textstyle{1 \over 2}} \sum_{\sigma \sigma'} \! \int \! d^{\scriptscriptstyle D}r \, d^{\scriptscriptstyle D}r' \, \psi_\sigma^\dagger({\bf r}) \, \psi_{\sigma'}^\dagger({\bf r}') \, 
U({\bf r} \! - \! {\bf r}') \, \psi_{\sigma'}({\bf r}') \, \psi_{\sigma}({\bf r}), 
\end{eqnarray*}
where ${\bf \Pi} \equiv {\bf p} + {\textstyle{e \over c}} {\bf A}$, and where $v_0({\bf r})$ is any single-particle potential energy, which may include a periodic lattice potential or disorder or both. Apart from an additive constant we can write $H$ as $H_0 + V$, where
\begin{equation}
H_0 \equiv \sum_\sigma \! \int \! d^{\scriptscriptstyle D}r \, \psi^\dagger_\sigma({\bf r}) \, \bigg[{\Pi^2 \over 2m} + v({\bf r}) - \mu \bigg] \psi_\sigma ({\bf r})
\label{H0}
\end{equation}
and
\begin{equation}
V \equiv {\textstyle{1 \over 2}} \int \! d^{\scriptscriptstyle D}r \, d^{\scriptscriptstyle D} r' \, \delta n({\bf r}) \, U({\bf r}-{\bf r}') \, \delta n({\bf r}').
\label{V}
\end{equation}
$H_0$ is the Hamiltonian in the Hartree approximation. The single-particle potential $v({\bf r})$ includes the Hartree interaction with the self-consistent density $n_0({\bf r})$,
\begin{equation*}
v({\bf r}) = v_0({\bf r}) + \int \! d^{\scriptscriptstyle D}r' \, U({\bf r} \! - \! {\bf r}') \, n_0({\bf r}'),
\end{equation*}
and the chemical potential has been shifted by $-U(0)/2$. In a translationally invariant system the equilibrium density is unaffected by interactions, but in a disordered or inhomogeneous system it will be necessary to distinguish between the approximate Hartree and the exact density distributions. $V$ is written in terms of the density fluctuation
$\delta n({\bf r}) \equiv \sum_\sigma \psi_\sigma^\dagger({\bf r})  \psi_\sigma({\bf r}) - n_0({\bf r}).$

The Euclidean propagator (\ref{G definition}) can be written in the interaction representation with respect to $H_0$ as
\begin{eqnarray}
G({\bf r}_{\rm f} \sigma_{\rm f}, {\bf r}_{\rm i} \sigma_{\rm i},\tau_0) \! &=& \!  - {\langle T \psi_{\sigma_{\rm f}}({\bf r}_{\rm f} ,\tau_0) {\bar \psi}_{\sigma_{\rm i}}
({\bf r}_{\rm i} ,0) e^{- \int_0^\beta d \tau \, V(\tau)} \rangle_0 \over \langle T e^{- \int_0^\beta d \tau \, V(\tau)} \rangle_0 } \nonumber \\
&=& {\cal N} \! \int \! D\mu[{\phi}] \ g({\bf r}_{\rm f} \sigma_{\rm f}, {\bf r}_{\rm i} \sigma_{\rm i},\tau_0 | \phi),
\label{transformed G}
\end{eqnarray}
where
\begin{equation}
D\mu[{\phi}] \equiv \frac{D\phi \, e^{-{1 \over 2} \int \phi U^{-1} \phi}}{ \int \! D\phi \ e^{-{1 \over 2} \int \phi U^{-1} \phi}} \ \ \ {\rm and} \ \ \ \int \! D\mu[{\phi}] = 1.
\end{equation}
Here  ${\cal N} \! \equiv \! \langle T \exp(-\int_0^\beta d \tau \, V) \rangle_0^{-1}$ is a $\tau_0$-independent constant, and
\begin{eqnarray}
&&g({\bf r}_{\rm f} \sigma_{\rm f}, {\bf r}_{\rm i} \sigma_{\rm i},\tau_0 | \phi) \nonumber \\
& {\hskip -0.2in} \equiv& {\hskip -0.1in} - \big\langle T \psi_{\sigma_{\rm f}}({\bf r}_{\rm f},\tau_0)  {\bar \psi}_{ \sigma_{\rm i}}({\bf r}_{\rm i},0) \, e^{i \int_0^\beta \! d \tau \int \! d^{\scriptscriptstyle D}r \, \phi({\bf r},\tau) \, \delta n({\bf r},\tau) } \big\rangle_0 {\hskip 0.20in}
\label{g correlation function definition}
\end{eqnarray}
is a correlation function describing noninteracting electrons in the presence of a purely {\it imaginary} scalar potential $-i \phi({\bf r},\tau)$. In the following we always assume that $\tau_0 \ge 0$.

Next we deform the contour of the functional integral by making the substitution $\phi \rightarrow i\phi_{\rm xr} + \phi,$ leading to\cite{contournote} 
\begin{eqnarray}
&&G({\bf r}_{\rm f} \sigma_{\rm f}, {\bf r}_{\rm i} \sigma_{\rm i},\tau_0) = {\cal N} e^{{1 \over 2} \! \int \! \phi_{\rm xr} U^{-1} \phi_{\rm xr}} \nonumber \\
& {\hskip -0.3in} \times & {\hskip -0.2in} \int \! D \mu[{\phi}] \, e^{-i \! \int \! \phi U^{-1} \phi_{\rm xr}} \,  g({\bf r}_{\rm f} \sigma_{\rm f}, {\bf r}_{\rm i} \sigma_{\rm i},\tau_0 | i\phi_{\rm xr} + \phi).  {\hskip 0.2in} 
\label{shifted functional integral}
\end{eqnarray}
This is an exact representation for the interacting Green's function, where $\phi$ now describes the {\it fluctuations} of the Hubbard-Stratonovich field about $\phi_{\rm xr}$.

\section{DOS IN THE X-RAY EDGE LIMIT}\label{xray limit section}

As explained above, our method involves identifying a certain field configuration $\phi_{\rm xr}$ that would be the potential produced by the tunneling particle if it had an infinite mass. Apart from a sign change, this potential is the same as that caused by a localized hole in the valence band of an optically excited metal. As is well known from work on the x-ray edge singularities, such fields cause an infrared catastrophe in the screening response. Fluctuations about $\phi_{\rm xr}$ account for the recoil of a real, finite-mass tunneling electron. 

In this section we introduce an approximation where these fluctuations are entirely neglected, which we shall refer to as the extreme x-ray edge limit. The approximation can itself be implemented in two different ways, perturbatively in the sense of Mahan\cite{MahanPR67} and exactly in the sense of Nozi\`{e}res and De Dominicis.\cite{Nozieres&DeDominicisPR69} In Sec.~\ref{application section}, we apply the perturbative x-ray edge method to the 1D electron gas, including spin, and to the 2D spin-polarized Hall fluid. The x-ray edge method will be implemented exactly for these and other models in future publications. 

In the x-ray edge limit, we ignore fluctuations about $\phi_{\rm xr}$ in $g({\bf r}_{\rm f} \sigma_{\rm f}, {\bf r}_{\rm i} \sigma_{\rm i},\tau_0 | i\phi_{\rm xr} + \phi)$, approximating it by $g({\bf r}_{\rm f} \sigma_{\rm f}, {\bf r}_{\rm i} \sigma_{\rm i},\tau_0 | i\phi_{\rm xr}).$\cite{taylornote} Then we obtain, from Eq.~(\ref{shifted functional integral}), 
\begin{equation}
G({\bf r}_{\rm f} \sigma_{\rm f}, {\bf r}_{\rm i} \sigma_{\rm i},\tau_0) = {\cal N} g({\bf r}_{\rm f} \sigma_{\rm f}, {\bf r}_{\rm i} \sigma_{\rm i},\tau_0| i\phi_{\rm xr}),
\label{propagator in xray limit}
\end{equation}
where, according to (\ref{g correlation function definition}), 
\begin{eqnarray}
&&g({\bf r}_{\rm f} \sigma_{\rm f}, {\bf r}_{\rm i} \sigma_{\rm i},\tau_0 | i \phi_{\rm xr}) \nonumber \\
& {\hskip -0.2in} = & {\hskip -0.1in} - \langle T \psi_{\sigma_{\rm f}}({\bf r}_{\rm f},\tau_0)  {\bar \psi}_{ \sigma_{\rm i}}({\bf r}_{\rm i},0) \, e^{- \! \int \! \phi_{\rm xr} \, \delta n} \rangle_0 . {\hskip 0.20in}
\label{gxr}
\end{eqnarray}
Eqs.~(\ref{propagator in xray limit}) and (\ref{gxr}) define the interacting propagator in the x-ray edge limit. The local tunneling DOS at position ${\bf r}_0$ is obtained by setting ${\bf r}_{\rm i} = {\bf r}_{\rm f} = {\bf r}_0$ and $\sigma_{\rm i} = \sigma_{\rm f} = \sigma_0$, and summing over $\sigma_0$.

\begin{widetext}

\subsection{Infrared catastrophe and divergence of perturbation theory}

To establish the connection between Eqs.~(\ref{propagator in xray limit}) and (\ref{gxr}) and the x-ray edge problem, we first calculate the tunneling DOS at ${\bf r}_0$ by evaluating (\ref{gxr}) perturbatively in $\phi_{\rm xr}$,
\begin{eqnarray}
&&g({\bf r}_0 \sigma_0, {\bf r}_0 \sigma_0,\tau_0 | i \phi_{\rm xr}) = G_0({\bf r}_0 \sigma_0, {\bf r}_0 \sigma_0,\tau_0) +  \int d^{\scriptscriptstyle D}r_1 \, d\tau_1 \, 
\phi_{\rm xr}({\bf r}_1,\tau_1) \, G_0({\bf r}_0 \sigma_0, {\bf r}_1 \sigma_0,\tau_0-\tau_1) \, G_0({\bf r}_1 \sigma_0, {\bf r}_0 \sigma_0,\tau_1) \nonumber \\
&+& \int d^{\scriptscriptstyle D}r_1 \, d^{\scriptscriptstyle D}r_2 \, d\tau_1 \, d\tau_2 \, \phi_{\rm xr}({\bf r}_1,\tau_1) \, \phi_{\rm xr}({\bf r}_2,\tau_2) \,
\bigg[ G_0({\bf r}_0 \sigma_0, {\bf r}_1 \sigma_0,\tau_0-\tau_1) \, G_0({\bf r}_1 \sigma_0, {\bf r}_2 \sigma_0,\tau_1 - \tau_2) \, 
G_0({\bf r}_2 \sigma_0, {\bf r}_0 \sigma_0,\tau_2) \nonumber \\
&-&  {\textstyle{1 \over 2}} \, G_0({\bf r}_0 \sigma_0, {\bf r}_0 \sigma_0 ,\tau_0 )  \, \Pi_0({\bf r}_1, {\bf r}_2 ,\tau_1-\tau_2) \bigg] + O\big(\phi_{\rm xr}^3\big),
\label{perturbation series in phi xr}
\end{eqnarray}
Here $G_0({\bf r} \sigma ,{\bf r}' \sigma',\tau) \equiv - \langle T \psi_{\sigma}({\bf r} , \tau) {\bar \psi}_{\sigma'}({\bf r}',0) \rangle_0$ is the mean-field Green's function associated with $H_0$, and
\begin{equation*} 
\Pi_0({\bf r}, {\bf r}' ,\tau) \equiv -\langle T \delta n({\bf r},\tau) \, \delta n({\bf r}',0)\rangle_0
\end{equation*}
is the density-density correlation function, given by $\sum_\sigma G_0({\bf r} \sigma ,{\bf r}' \sigma,\tau) \, G_0({\bf r}' \sigma ,{\bf r} \sigma,-\tau).$ Now we use (\ref{DOS phi xr}), and assume the short-range interaction (\ref{short range interaction definition}). Then Eq.~(\ref{perturbation series in phi xr}) reduces to
\begin{eqnarray}
&&g({\bf r}_0 \sigma_0, {\bf r}_0 \sigma_0,\tau_0 | i \phi_{\rm xr}) = G_0({\bf r}_0 \sigma, {\bf r}_0 \sigma,\tau_0) + \lambda \int_0^{\tau_0} \! \! d\tau_1 \, 
G_0({\bf r}_0 \sigma_0, {\bf r}_0 \sigma_0,\tau_0-\tau_1) \, G_0({\bf r}_0 \sigma_0, {\bf r}_0 \sigma_0,\tau_1) \nonumber \\
&+&  \lambda^2  \int_0^{\tau_0} \! \! d\tau_1 \, d\tau_2 \, \bigg[ G_0({\bf r}_0 \sigma_0, {\bf r}_0 \sigma_0,\tau_0-\tau_1) \, G_0({\bf r}_0 \sigma_0, {\bf r}_0 \sigma_0,\tau_1 - \tau_2) \, G_0({\bf r}_0 \sigma_0, {\bf r}_0 \sigma_0,\tau_2) 
- {\textstyle{1 \over 2}} \, G_0({\bf r}_0 \sigma_0, {\bf r}_0 \sigma_0 ,\tau_0 )  \, \Pi_0({\bf r}_0, {\bf r}_0 ,\tau_1 - \tau_2) \bigg] \nonumber \\
&+& O\big(\lambda^3\big).
\label{perturbation series in lambda}
\end{eqnarray}

We evaluate (\ref{perturbation series in lambda}) for the 1D electron gas and the 2D Hall fluid in the large $\tau_0 \epsilon_{\rm F}$ limit using the low-energy propagators of Appendix \ref{noninteracting propagator appendix}. $\epsilon_{\rm F}$ is the Fermi energy. In the 1D electron gas case this leads to
\begin{equation}
g({\bf r}_0 \sigma_0, {\bf r}_0 \sigma_0,\tau_0 | i \phi_{\rm xr}) = G_0({\bf r}_0 \sigma_0, {\bf r}_0 \sigma_0,\tau_0) \bigg\lbrace 1 - 2 \lambda N_0 \ln(\tau_0 \epsilon_{\rm F})
+ \lambda^2 N_0^2 \bigg[ 2 \ln^2(\tau_0 \epsilon_{\rm F}) - 2 \ln(\tau_0 \epsilon_{\rm F}) + {\pi \over 2} \tau_0 \epsilon_{\rm F} \bigg] + \cdots \bigg\rbrace,
\label{perturbation series for 1DEG}
\end{equation}
where we have used the asymptotic results given in Appendix \ref{integrals appendix} and have kept only the corrections through order $\lambda^2$ that diverge in the large $\tau_0$ limit. Here $N_0$ is the noninteracting DOS per spin component at $\epsilon_{\rm F}.$\cite{DOSnote} These divergences are caused by the infrared catastrophe and the associated breakdown of perturbation theory at low energies. Similarly, for the Hall fluid we find
\begin{equation}
g({\bf r}_0 \sigma_0, {\bf r}_0 \sigma_0,\tau_0 | i \phi_{\rm xr}) = G_0({\bf r}_0 \sigma_0, {\bf r}_0 \sigma_0,\tau_0) \bigg\lbrace 1 
- \lambda \bigg(\frac{1-\nu}{2 \pi \ell^2}\bigg) \tau_0 + \frac{1}{2} \lambda^2 \bigg(\frac{1-\nu}{2 \pi \ell^2}\bigg)^{\! 2}  \tau_0^2  + \cdots \bigg\rbrace,
\label{perturbation series for LLL}
\end{equation}
where $\nu$ is the filling factor and $\ell$ is the magnetic length. The divergence in this case is stronger because of the infinite compressibility of the fractionally filled Landau level at mean field level.

\end{widetext}

\subsection{Perturbation series resummation}

A logarithmically divergent perturbation series similar to (\ref{perturbation series for 1DEG}) occurs in the x-ray edge problem, where it is known that qualitatively correct results are obtained by reorganizing the series into a second-order cumulant expansion. Here we will carry out such a resummation for both  (\ref{perturbation series for 1DEG}) and
(\ref{perturbation series for LLL}).

In the electron gas case this leads to
\begin{eqnarray}
g({\bf r}_0 \sigma_0, {\bf r}_0 \sigma_0,\tau_0 | i \phi_{\rm xr})  \! &\approx& \! G_0({\bf r}_0 \sigma_0, {\bf r}_0 \sigma_0,\tau_0) \nonumber \\
&\times& \! \bigg( \frac{1}{\tau_0 \epsilon_{\rm F}} \bigg)^{\! \alpha} e^{ {\pi \over 2} \lambda^2 N_0^2 \tau_0 \epsilon_{\rm F}} ,
\label{cumulant resummation 1DEG}
\end{eqnarray}
where
\begin{equation}
\alpha \equiv 2 \lambda N_0 + 2  \lambda^2 N_0^2.
\label{alpha exponent}
\end{equation}
The exponential factor in (\ref{cumulant resummation 1DEG}), after analytic continuation, produces a negative energy shift. However, this shift depends sensitively on the short-time regularization and is not reliably calculated with our method. $\alpha$ is equal to the x-ray absorption/emisssion edge exponent 
\begin{equation*}
2 (\varphi/\pi) + 2 (\varphi/\pi)^2
\end{equation*} 
of Nozi\`{e}res and De Dominicis\cite{Nozieres&DeDominicisPR69} (including spin) for a {\it repulsive} potential, with $\varphi \equiv \arctan (\pi \lambda N_0)$ the phase shift at $\epsilon_{\rm F}$, when expanded to order $\lambda^2$.

In the Hall case with $0 < \nu < 1$,
\begin{equation}
{\hskip -0.06in} g({\bf r}_0 \sigma_0, {\bf r}_0 \sigma_0,\tau_0 | i \phi_{\rm xr}) \approx \frac{\nu - 1}{2 \pi \ell^2} \, e^{\gamma (\nu-1)\tau_0},
\label{cumulant resummation for LLL}
\end{equation}
where
\begin{equation}
\gamma \equiv \frac{\lambda}{2 \pi \ell^2}
\end{equation}
is an interaction strength with dimensions of energy. This time the infrared catastrophe causes a positive energy shift and no transient relaxation. The energy shift in this case is predominantly determined by long-time dynamics and {\it is} physically meaningful.

\section{APPLICATIONS OF THE PERTURBATIVE X-RAY EDGE LIMIT}\label{application section}

The tunneling DOS is obtained by analytic continuation. We take the zero of energy to be the mean-field Fermi energy. 

\subsection{1D electron gas}

For the 1D electron gas case we find a power-law
\begin{equation}
N(\epsilon) = {\rm const} \! \times \! \epsilon^\alpha,
\label{electron gas xray DOS}
\end{equation}
where the exponent $\alpha$ is given in (\ref{alpha exponent}). In (\ref{electron gas xray DOS}) we have neglected the energy shift appearing in (\ref{cumulant resummation 1DEG}).

The power-law DOS we obtain is qualitatively correct, although the value of the exponent certainly is not. The exponent will be modified by carrying out the x-ray edge limit exactly, and also presumably by including fluctuations about $\phi_{\rm xr}$. However, the fact that we recover the generic algebraic DOS of the Tomonaga-Luttinger liquid phase is at least consistent with our assertion that the algebraic DOS in 1D metals is caused by the infrared catastrophe.

We note that the x-ray edge approximation actually predicts a power-law DOS in clean 2D and 3D electron systems in zero field as well. However, in those cases there is no reason to expect the x-ray edge limit to be relevant: In those cases the electron recoil and hence fluctuations about $\phi_{\rm xr}$ are large.

\subsection{2D Hall fluid}

For the 2D Hall fluid with $0 < \nu < 1$ we obtain
\begin{equation}
N(\epsilon) = {\rm const} \! \times \! \delta\big(\epsilon - [1-\nu]\gamma \big).
\label{hall fluid xray DOS}
\end{equation}
The lowest Landau is moved up in energy by an amount
\begin{equation}
\lambda \, \frac{1-\nu}{2 \pi \ell^2}.
\label{gap size}
\end{equation}

The DOS (\ref{hall fluid xray DOS}) is also qualitatively correct, in the following sense: The actual DOS in this system is thought to be a broadened peak at an energy of about $e^2/\kappa \ell$, with $\kappa$ the dielectric constant of the host semiconductor, producing a pseudogap at $\epsilon_{\rm F}$. Of course the magnetic field dependence of the peak positions are different, but the physical system has a screened Coulomb interaction, and and here we obtain a hard ``gap" of size (\ref{gap size}) at $\epsilon_{\rm F}$. We speculate that the cumulant expansion gives the exact x-ray edge result for this model, but that fluctuations about $\phi_{\rm xr}$ will broaden the peak in (\ref{hall fluid xray DOS}).

\section{DISCUSSION}\label{discussion section}

The results and (\ref{electron gas xray DOS}) and (\ref{hall fluid xray DOS}) are consistent with our claim that the DOS anomalies in the 1D electron gas and the 2D spin-polarized Hall fluid are caused by an infrared catastrophe, similar to that responsible for the singular x-ray spectra of metals, the orthogonality catastrophe, and the Kondo effect. (The models we considered are of course tractable by existing specialized methods.) In addition to providing a common explanation for a variety of tunneling anomalies, our method may provide a means of calculating the DOS in other strongly correlated and low-dimensional systems, such as at the edge of the sharply confined Hall fluid investigated experimentally by Grayson {\it et al.},\cite{GraysonPRL98} Chang {\it et al.},\cite{ChangPRL01} and by Hilke {\it et al.},\cite{HilkePRL01} where existing theoretical methods fail.

Anderson has gone even further and proposed that this same infrared catastrophe causes a complete breakdown of Fermi liquid theory in 2D electron systems, such as the Hubbard model, in zero field.\cite{AndersonPRL90a,AndersonPRL90b,AndersonPRL93} We are at present unable to address this question with the methods described here, which assume that the effects of fluctuations about $\phi_{\rm xr}$ are small.

Our method can be applied, in principle, to realistic tunneling geometries, with or without disorder, and with Coulomb interaction.

\appendix

\section{LOW-ENERGY NONINTERACTING PROPAGATOR}\label{noninteracting propagator appendix}

For a noninteracting electron system with single-particle states $\phi_\alpha({\bf r})$ and spectrum $\epsilon_\alpha,$ the imaginary-time Green's function defined in Eq.~(\ref{G definition}) is given by
\begin{eqnarray}
&& G_0({\bf r} \sigma ,{\bf r}' \sigma',\tau) = \delta_{\sigma \sigma'} \sum_\alpha \phi_\alpha({\bf r}) \phi^*_{\alpha}({\bf r}') e^{-(\epsilon_\alpha - \mu)\tau} \nonumber \\
&\times& \! \! \bigg( \big[ n_{\rm F}(\epsilon_\alpha - \mu)-1 \big] \ \! \Theta(\tau) + n_{\rm F}(\epsilon_\alpha - \mu) \ \! \Theta(-\tau) \bigg), \ \ \ \ \ \ \ \ \ \ 
\label{general G0}
\end{eqnarray}
where $n_{\rm F}(\epsilon) \equiv (e^{\beta \epsilon} + 1)^{-1}$ is the Fermi distribution function. In this appendix we shall evaluate the diagonal components $G_0({\bf r} \sigma ,{\bf r} \sigma,\tau)$ for two models in the large $|\tau|$, asymptotic limit.

\subsection{1D electron gas}

The first model is a translationally invariant electron gas at zero temperature in 1D (the derivation we give is actually valid for any dimension $D$). In the limit of large $ \epsilon_{\rm F} |\tau|$ we obtain
\begin{equation}
G_0({\bf r} \sigma ,{\bf r} \sigma,\tau) \rightarrow - \frac{N_0}{\tau},
\label{electron gas asymptotic G0}
\end{equation}
where $N_0$ is the noninteracting DOS per spin component at the Fermi energy $\epsilon_{\rm F}.$\cite{DOSnote} It will be necessary to regularize the unphysical short-time behavior in Eq.~(\ref{electron gas asymptotic G0}). The precise method of regularization will not affect our final results of interest, such as exponents, which are determined by the long-time behavior. We will take the regularized asymptotic propagator to be
\begin{equation}
G_0({\bf r} \sigma ,{\bf r} \sigma,\tau) \approx -  \, {\rm Re} \, \frac{N_0}{\tau + i / \epsilon_{\rm F}} .
\label{electron gas regularized asymptotic G0}
\end{equation}
When possible, we will let the short-time cutoff in (\ref{electron gas regularized asymptotic G0}) approach zero, in which case (\ref{electron gas regularized asymptotic G0}) simplifies to 
\begin{equation}
G_0({\bf r} \sigma ,{\bf r} \sigma,\tau) \approx -  \, {\rm P} \, \frac{N_0}{\tau} ,
\label{electron gas P regularized G0}
\end{equation}
where P denotes the principal part. The results  (\ref{electron gas regularized asymptotic G0}) and (\ref{electron gas P regularized G0}) are valid for any spatial dimension $D$; the only $D$ dependence appears in the value of $N_0$ for a given $\epsilon_{\rm F}.$

\subsection{2D Hall fluid}

The second model we consider is a 2D spin-polarized electron gas in the lowest Landau level at zero temperature with filling factor $\nu$. In the gauge ${\bf A} = B x {\bf e}_y$, 
\begin{equation}
\phi_{nk}({\bf r}) = c_{nk} \ \! e^{-iky}  \ \! e^{-{1 \over 2}(x/\ell - k \ell )^2} \ \! H_n(x/\ell - k \ell),
\label{Landau orbitals}
\end{equation}
where $c_{nk} \! \equiv \! ( 2^n n! \pi^{1 \over 2} \ell L )^{-{1\over 2}}.$ Here $\ell \equiv \sqrt{\hbar c/eB}$ is the magnetic length and $L$ is the system size in the $y$ direction. The spectrum is $\epsilon_n = \hbar \omega_{\rm c} (n+{\textstyle{1 \over 2}}),$ with $\omega_{\rm c} \equiv eB/m^*c$ the cyclotron frequency ($m^*$ is the band mass). In the lowest Landau level $0 < \nu \le 1$,
\begin{equation}
\phi_k =  (\pi^{1 \over 2} \ell L)^{-{1 \over 2}} \ \!  e^{-iky} \ \! e^{-{1 \over 2}(x/\ell - k \ell )^2}. 
\label{LLL orbitals}
\end{equation}
At long times $\tau \gg \omega_{\rm c}^{-1}$ we obtain
\begin{equation}
G_0({\bf r},{\bf r},\tau) = \frac{\nu -  \Theta(\tau)}{2 \pi \ell^2} .
\label{LLL asymptotic G0}
\end{equation}

\section{ASYMPTOTIC EVALUATION OF TIME INTEGRALS}\label{integrals appendix}

Here we note the asymptotics used to obtain (\ref{perturbation series for 1DEG}): 
\begin{eqnarray*}
\int_0^{\tau_0} \! d\tau \  {\rm Re} \bigg( \frac{1}{\tau_0 - \tau + i/\epsilon_{\rm F}} \bigg) \, {\rm Re} \bigg( \frac{1}{\tau + i/\epsilon_{\rm F}} \bigg) 
\approx  \frac{2}{\tau_0}  \ln (\tau_0 \epsilon_{\rm F}),
\end{eqnarray*}

\begin{eqnarray*}
&&\int_0^{\tau_0} \! d\tau \, d\tau' \  {\rm Re} \bigg( \frac{1}{\tau_0 - \tau + i/\epsilon_{\rm F}} \bigg) \, {\rm Re} \bigg( \frac{1}{\tau - \tau' + i/\epsilon_{\rm F}} \bigg) 
\nonumber \\
& \times& {\rm Re} \bigg( \frac{1}{\tau' + i/\epsilon_{\rm F}} \bigg) \approx  \frac{2}{\tau_0}  \ln^2 (\tau_0 \epsilon_{\rm F}),
\end{eqnarray*}
and
\begin{eqnarray*}
\int_0^{\tau_0} \! d\tau \, d\tau' \, \bigg[ {\rm Re} \bigg( \frac{1}{\tau - \tau' + i/\epsilon_{\rm F}} \bigg)\bigg]^2 \approx  {\pi \over 2} \, \tau_0 \epsilon_{\rm F} - 2 \ln (\tau_0 \epsilon_{\rm F}).
\end{eqnarray*}
In these expressions we have retained all terms, including subdominant contributions, that diverge in the $\tau_0 \rightarrow \infty$ limit.

\acknowledgments

This work was supported by the National Science Foundation under CAREER Grant No.~DMR-0093217. It is a pleasure to thank Phil Anderson, Claudio Chamon, Matthew Grayson, Dmitri Khveshchenko, Yong Baek Kim, Allan MacDonald, Gerry Mahan, David Thouless, Shan-Ho Tsai, Giovanni Vignale, and Ziqiang Wang for useful discussions. 

\bibliography{/Users/mgeller/Papers/bibliographies/MRGhall,/Users/mgeller/Papers/bibliographies/MRGmanybody,/Users/mgeller/Papers/bibliographies/MRGpre,/Users/mgeller/Papers/bibliographies/MRGbooks,/Users/mgeller/Papers/bibliographies/MRGgroup,xraynotes}

\begin{thebibliography}{66}
\bibitem[{tay()}]{taylornote}
\bibinfo{note}{This approximation corresponds to regarding $\phi$ as small,
  Taylor expanding about $i \phi_{\rm xr}$, and keeping only the zeroth-order
  term in the series. However, it is unclear whether such a functional Taylor
  expansion is meaningful here.}
\bibitem[{DOS()}]{DOSnote}
\bibinfo{note}{For a free electron gas with parabolic dispersion in $D$
  dimensions, $$ N_0 = \frac{D {\cal S}_{\scriptscriptstyle D} \epsilon_{\rm
  F}^{D-1}}{2 (\pi \hbar v_{\rm F})^D}, $$ where ${\cal S}_{\scriptscriptstyle
  D}$ is the $D$-dimensional unit sphere volume and $v_{\rm F}$ is the Fermi
  velocity.}
\end{thebibliography}

\end{document}